\documentclass{article}
\relax
\usepackage{ijcai19}  
\usepackage{setspace}
\usepackage{xspace}
\usepackage{enumitem}
\usepackage{enumerate}

\usepackage[utf8]{inputenc} 
\usepackage[T1]{fontenc}    
\usepackage{url}            
\usepackage{booktabs}       
\usepackage{amsfonts}       
\usepackage{nicefrac}       
\usepackage{microtype}      
\usepackage{xspace}
\usepackage{times}
\usepackage{helvet}
\usepackage{courier}
\usepackage{algorithm}
\usepackage{algorithmic}
\usepackage{amsthm}
\usepackage{bm}

\usepackage{makeidx}  
\usepackage{amsmath}
\usepackage{epsfig}
\usepackage{enumerate}
\usepackage{subfigure}
\usepackage{makecell}
\usepackage{multirow}
\usepackage{epstopdf}
\usepackage{graphicx}
\usepackage{amsfonts}

\usepackage{caption}
\usepackage[colorlinks=true, allcolors=blue]{hyperref}
\usepackage [autostyle, english = american]{csquotes}
\MakeOuterQuote{"}

\newcommand{\mname}{\texttt{MINA}\xspace}

\title{\mname: Multilevel Knowledge-Guided Attention for Modeling \\ Electrocardiography Signals}

\author{
Shenda Hong$^{1,2,5}$\and
Cao Xiao$^3$\and
Tengfei Ma$^4$\and
Hongyan Li$^{1,2}$\And
Jimeng Sun$^5$\\
\affiliations
$^1$School of Electronics Engineering and Computer Science, Peking University, China\\
$^2$Key Laboratory of Machine Perception (Ministry of Education), Peking University, China\\
$^3$Analytics Center of Excellence, IQVIA, USA\\
$^4$IBM Research, USA\\
$^5$Department of Computational Science and Engineering, Georgia Institute of Technology, USA\\
\emails
hongshenda@pku.edu.cn,
cao.xiao@iqvia.com,
Tengfei.Ma1@ibm.com,
lihy@cis.pku.edu.cn,
jsun@cc.gatech.edu
}

\hypersetup{draft}

\begin{document}

\newcounter{sol} 
\setcounter{sol}{1} 

\maketitle

\begin{abstract}
Electrocardiography (ECG) signals are commonly used to diagnose various cardiac abnormalities. Recently, deep learning models showed initial success on modeling ECG data, however they are mostly black-box, thus lack interpretability needed for clinical usage. In this work, we propose MultIlevel kNowledge-guided Attention networks (\mname) that predict heart diseases from ECG signals with intuitive explanation aligned with medical knowledge. By extracting multilevel (beat-, rhythm- and frequency-level) domain knowledge features separately, \mname combines the medical knowledge and ECG data via a multilevel attention model, making the learned models highly interpretable. Our experiments showed \mname achieved PR-AUC $0.9436$ (outperforming the best baseline by $5.51\%$) in real world ECG dataset. Finally, \mname also demonstrated robust performance and strong interpretability against signal distortion and noise contamination.
\end{abstract}

\section{Introduction}
\label{sec:introduction}

Heart diseases are among the leading causes of death of the world~\cite{benjamin2018heart}. The routine monitoring of physiological signals is deemed important in heart disease prevention. Among existing monitoring technologies, electrocardiography (ECG) is a commonly used non-invasive and convenient diagnostic tool that records physiological activities of heart over a period of time. 
Deciphering ECG signals can help detect many heart diseases such as atrial fibrillation (AF), myocardial infarction (MI), and heart failure (HF)~\cite{jama_af,yanowitz2012introduction}.

\begin{figure}[t]
\centering
\includegraphics[width=0.48\textwidth]{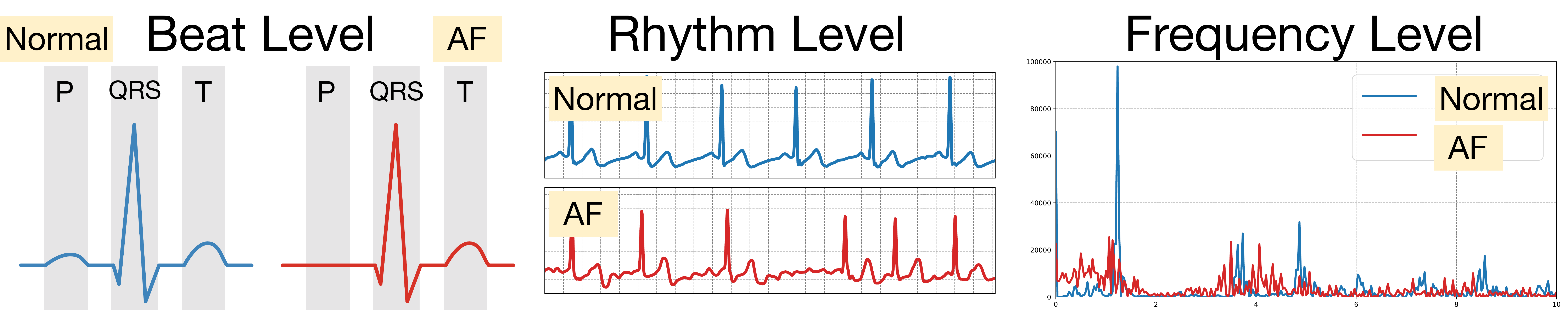}
\caption{Normal ECG signal and Abnormal ECG signal show different patterns across different levels.}
\vspace{-0.25cm}
\label{fig:example_intro}
\end{figure}

An example of real world ECG signal is shown in Figure\ref{fig:example_intro}. ECG signals from cases and controls of heart diseases show different patterns at 1) \textit{beat level}, 2) \textit{rhythm level}, and 3) \textit{frequency level}, each representing different anomalous activities of the heart. For example, beat level morphology such as P wave (atrial depolarization) and QRS complex (ventricular depolarization) can reflect conditions related to heart electric conduction. Rhythm level patterns capture rhythm features across beats and reflect cardiac arrhythmia conditions (abnormal heart rhythms). Frequency level is about frequency variations and sheds light on the diagnosis of ventricular flutter and ventricular fibrillation. Learning these patterns to support diagnoses has been an important research area in ECG analysis~\cite{roopa2017survey,expert_1,expert_3,tateno2001automatic}.

In real clinical settings, in addition to the demand of an accurate classification, the interpretability of the results is equally important ~\cite{tsai2003computer}. Cardiologists need to provide both diagnosis and detailed explanations to support diagnosis~\cite{std}. Also,  many heart diseases do not pose abnormal ECG diagram constantly~\cite{benjamin2018heart,yanowitz2012introduction}, especially during the early stage of the diseases. Therefore, interpretability of the results, particularly highlighting diagnosis related parts of the data, is crucial for early diagnosis and better clinical decisions. 

Traditional machine learning methods either learn time domain patterns including beat level~\cite{ladavich2015rate,purerfellner2014p} and rhythm level~\cite{huang2011novel},
or extract frequency patterns using signal processing techniques such as discrete wavelet transform ~\cite{garcia2016application}. However, time domain approaches are easily affected by noise or signal distortion~\cite{RODRIGUEZ2015261}; while frequency domain methods cannot model rare events or some temporal dynamics that occur in time domain. Besides, they all require laborious feature engineering, and their performance also relies on the quality of the constructed features.

Recently, deep learning models showed initial success in modeling ECG data. Convolutional neural networks (CNN) were used to learn beat level patterns~\cite{tbe,2017arXiv170701836R,hannun2019cardiologist}. Recurrent neural networks (RNN) are suitable for capturing rhythm features~\cite{schwab2017beat,hong2017encase,zihlmann2017convolutional}. Moreover, attention mechanism is employed to extract interpretable rhythm features ~\cite{schwab2017beat}. Despite their progress, these models were either black-box or only highlighted one aspect of patterns (such as rhythm features as in~\cite{schwab2017beat}), thus lack the comprehensive interpretability of the results for real clinical usage.

In this work, we propose {\it M}ult{\it I}level k{\it N}owledge-guide {\it A}ttention model (\mname) to learn and integrate different levels of features from ECG which are aligned with clinical knowledge. For each level \mname extracts level-specific domain knowledge features and uses them to guide the attention, including beat morphology knowledge that guides attentive CNN and rhythm knowledge that guides attentive RNN. \mname also performs attention fusion across time- and frequency domains. 
We proposed new evaluation approaches by interfering ECG signals with noise and signal distortion. We evaluated  interpretability and robustness of the model by tracking intermediate reactions across layers from multilevel attentions to the final predictions. 

Experimental results show \mname can correctly identify critical beat location, significant rhythm variation, important frequency component and remain robust in prediction under signal distortion or noise contamination.
Tested on the atrial fibrillation prediction, \mname achieved PR-AUC $0.9436$ (outperforming the best baseline by $5.51\%$). 
Finally, \mname also showed strong result interpretability and more robust performance than baselines.

\section{Related Work}
\label{sec:related_work}

Traditional methods include time domain methods such as beat level methods~\cite{ladavich2015rate,purerfellner2014p} and rhythm level ones ~\cite{tateno2001automatic,huang2011novel,oster2015impact}, both depending on segmentation by detecting QRS complex. However, time domain methods rely on the accuracy of QRS detection, thus are easily affected by noise or signal distortion.  Frequency domain approaches, on the other hand, cannot model rare events and other time-domain patterns and thus lack interpretability. Moreover, both types of features are subjective.

Recently, deep neural networks (DNNs) have been used in ECG diagnosis~\cite{tbe,2017arXiv170701836R,hannun2019cardiologist,zihlmann2017convolutional,hong2017encase,schwab2017beat}. Many of them have demonstrated state-of-the-art performance due to their ability in extracting effective features~\cite{2017arXiv170701836R,hong2017encase}. Some of them build an end-to-end classifier ~\cite{tbe,2017arXiv170701836R,zihlmann2017convolutional}, others build a mixture model which combines traditional feature engineering methods and deep models ~\cite{hong2017encase,schwab2017beat,hong2019}. However, existing deep models are insufficient in three aspects. First, they neglect the characteristics of ECG signals when design model architecture, namely, beat morphological, rhythm variations. Second, they only analyze ECG signals in time domain. Last, they are "black-box" and thus not interpretable. In real world medical applications, interpretability is critical for clinicians to accept machine recommendations and implement intervention. 

\section{Method}
\label{sec:method}

\begin{figure}[t]
\centering
\includegraphics[width=0.48\textwidth]{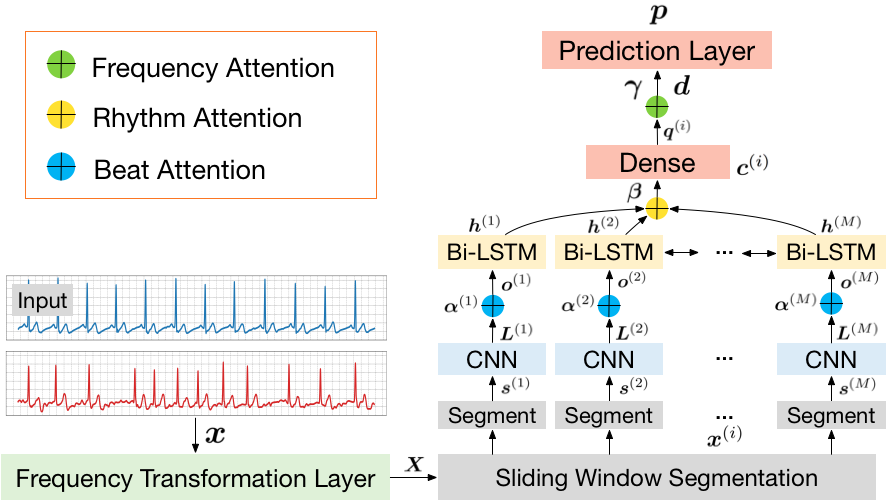}
\vspace{-0.5cm}
\caption{\mname takes raw ECG signals as input and outputs probabilities of disease onset. \mname used knowledge-guided attention to learn informative beat-, rhythm-, and frequency level patterns, and then performs attentive signal fusion for improved prediction.}
\vspace{-0.25cm}
\label{fig:net}
\end{figure}

In this section, we will introduce the model design of \mname. 
Section \ref{sec:notation} provides an overview and introduces all notations. Section~\ref{sec:framework} describes the basic framework, including each layer of \mname. Section \ref{sec:knowledge} proposes our new attention mechanism which is integrated in \mname. Section~\ref{sec:interprete} describes how we evaluate interpretability and robustness. Figure\ref{fig:net} depicts the architecture of \mname. 

\subsection{Overview of \mname}
\label{sec:notation}
Here we briefly describe the framework and introduce notations used in this paper. Assume we are given a single lead ECG signal $\bm{x} \in \mathbb{R}^n $ and use it to predict class probability.  We firstly transform it into multi-channel signals with $F$ channels across different frequency bands where $i$th signal is denoted as $\bm{x}^{(i)} \in \mathbb{R}^n $. We then split each  $\bm{x}^{(i)}$ into $M$ segments $\bm{s}^{(k)}$. Next we apply CNN and RNN consecutively on $\bm{s}^{(k)}$ to obtain beat level attention $\bm{o}^{(k)}, 1 \leq k \leq M$ and rhythm level attention $\bm{c}^{(i)}$. This follows by a fully connected layer that transforms $\bm{c}^{(i)}$ into $\bm{q}^{(i)}$. We then take weighted average to integrate $\bm{Q}=[\bm{q}^{(1)},...,\bm{q}^{(F)}]$ across all channels to output frequency attention $\bm{d}$, which will be used in prediction. To improve model accuracy and interpretability, we propose a \textit{knowledge guided attention} to learn attention vectors from beat-, rhythm-, and frequency levels, denoted as $\bm{\alpha}$, $\bm{\beta}$, and $\bm{\gamma}$ respectively.
More details will be described in Section~\ref{sec:framework}. The notations are summarized in Table~\ref{tb:notations}. Detailed configurations of \mname are shown in the Experiments section. 

\begin{table}[t]
\centering
\resizebox{0.48\textwidth}{!}{
\begin{tabular}{l|l}
\hline
\textbf{Notation} & \textbf{Definition}  \\
\hline
$C$,$F$& \# of classes, \# of frequency channels\\
$n$,$M$,$T$ & ECG length, \# of segments, segment length\\ 
$\bm{x} \in \mathbb{R}^n $ & Original ECG signal \\
$\bm{X} \in \mathbb{R}^{F \times n} $, $\bm{x}^{(i)} \in \mathbb{R}^n $ & Signals after transformation, $i$th signal \\
$\bm{s} \in \mathbb{R}^{T}$ & Segment of ECG with length $T$ \\ 
$\bm{S} \in \mathbb{R}^{M \times T}$, $\bm{s^{(k)}} \in \mathbb{R}^{T}$ & $M$ segments, $k$th segment\\ 
\hline
$\bm{L} \in \mathbb{R}^{K \times N}$ & CNN layer output  \\ 
$\bm{l}^{(j)} \in \mathbb{R}^{K}, \bm{L}^{(k)} \in \mathbb{R}^{K \times N}$ & $j$th column in $\bm{L}$, $k$th segment output  \\ 
$\bm{o} \in \mathbb{R}^{K}$ & Output of beat level attention \\
$\bm{O} \in \mathbb{R}^{K \times M}$, $\bm{o}^{(k)} \in \mathbb{R}^{K}$  & Output of beat attention of $M$ segments, $k$th output \\ \hline
$\bm{H} \in \mathbb{R}^{J \times M}$, $\bm{h}^{(k)} \in \mathbb{R}^{J}$ & Bi-LSTM layer output, $k$th column in $\bm{H}$ \\
$\bm{c} \in \mathbb{R}^{J}$ & Output of rhythm level attention \\ 
$\bm{C} \in \mathbb{R}^{J \times F}$, $\bm{c}^{(i)} \in \mathbb{R}^{J}$ & Output of rhythm attention of $F$ channels, $i$th output  \\ \hline
$\bm{W_c} \in \mathbb{R}^{J \times H}, \bm{b_c} \in \mathbb{R}^{H}$ & Weight and bias in fully connected layer\\
$\bm{Q} \in \mathbb{R}^{H \times F}$, $\bm{q}^{(i)} \in \mathbb{R}^{H}$ & Fully connected layer output, $i$th column in $\bm{Q}$ \\
$\bm{d} \in \mathbb{R}^{H}$ & Output of freq. level attention \\ \hline
$\bm{W} \in \mathbb{R}^{H \times C}, \bm{b} \in \mathbb{R}^{C}$ & Weight matrix and bias vector in prediction layer\\
$\bm{p}$, $\bm{p}_c$ & Predicted probability, $c$th value \\
$\bm{w}$, $\bm{w}_c$ & Class weight, $c$th value \\
$\bm{z}$, $\bm{z}_c$ & One-hot label, $c$th value \\
\hline
$\bm{\alpha} \in \mathbb{R}^{N}$ & Beat attention weights \\
$\bm{\alpha}_j \in \mathbb{R}$, $\bm{\alpha}^{(k)} \in \mathbb{R}^{N}$ & $j$th value in  $\bm{\alpha}$, segment $k$ attention\\
$\bm{\beta} \in \mathbb{R}^{M}$, $\bm{\beta}_k \in \mathbb{R}$ & Rhythm level attention weights, $k$th value in $\bm{\beta}$ \\
$\bm{\gamma} \in \mathbb{R}^{F}$, $\bm{\gamma}_i \in \mathbb{R}$ & Frequency level attention weights,  $i$th value in $\bm{\gamma}$\\
$\bm{K}_{*} \in \mathbb{R}^{E_{*} \times N}$ & Knowledge feature ($*$ can be $\alpha$, $\beta$ or $\gamma$) \\
$\bm{W}_{*} \in \mathbb{R}^{(K+E_{*}) \times D_{*}}$ & $1$st layer attention weights ($*$ can be $\alpha$, $\beta$ or $\gamma$)\\
$\bm{b}_{*} \in \mathbb{R}^{D_{*}}$ & $1$st layer attention biases ($*$ can be $\alpha$, $\beta$ or $\gamma$)\\
$\bm{V}_{*} \in \mathbb{R}^{D_{*} \times 1}$ & $2$nd layer attention weights ($*$ can be $\alpha$, $\beta$ or $\gamma$)\\
\hline
$\mathcal{D}$ & Function of standard deviation \\
$\mathcal{F}$ & Function of power spectral density \\
\hline
$\bm{x'}, \bm{\alpha'}, \bm{\beta'}, \bm{\gamma'}, \bm{p'}$ & Interfered signals, attention weights and predictions \\
\hline
\end{tabular}
}
\caption{Notations for \mname}
\label{tb:notations}
\end{table}

\subsection{Description of \mname}
\label{sec:framework}
\subsubsection{Signal Transformation and Segmentation}
In order to utilize the frequency-domain information, we employ an efficient strategy by decomposing original ECG signals into different frequency bands (where each band is regarded as a channel). Then we can concurrently model signals of each channel.

Specifically, we propose a new time-frequency transformation layer to transform a \textit{single lead} ECG signal into multi-channel ones. Here we use Finite Impulse Response (FIR) bandpass filter~\cite{Oppenheim:1996:SAS:248702} to transform  \textit{single lead} ECG signal $\bm{x}$ into $F$ multi-channel ECG signals  $\bm{X} = [\bm{x}^{(1)}, \bm{x}^{(2)}, ..., \bm{x}^{(F)}]$. 

Then for each channel, we split $\bm{x}^{(i)}\in \mathbb{R}^n$ into a sequence of $M$ equal length segments. Unlike previous deep models ~\cite{schwab2017beat,tbe} that perform segmentation using QRS complex detection, which is easily affected by signal quality, we simply use sliding window segmentation. By cutting each of $i$th segment is indexed by $(i-1)\times T$ and $i\times T - 1$, we receive $M$ equal length segments $\bm{s} \in \mathbb{R}^T$ (without the loss of generality, we assume that $n=M*K$, otherwise we can cut off last remain part which is shorter than $T$). In general, segment length $T$ needs to be shorter than the length of one heart beat, so that we can extract patterns in beat level. 
Detailed configurations can be found in Implementation Details section.

\subsubsection{Beat Level Attentive Convolutional Layer}
For beat level patterns, we mainly consider the abnormal wave shapes or edges. To locate them from signals, we design an attentive convolutional layer. 
Formally, given $M$ segments $\bm{s} \in \mathbb{R}^{T}$, we perform 1-D convolution on each of them and output convolved features: $\bm{L} =Conv(\bm{s})$, $\bm{L} \in \mathbb{R}^{K \times N}$, $K$ is the number of filters, $N$ is the output length of segments after convolution, which is determined by hyperparameters like stride of CNN. $Conv$ operations are shared weights of $M$ segments. Then instead of traditional global average pooling which treats all features homogeneously, we propose a \textit{knowledge-guided attention} to aggregate these features and get beat level attention $\bm{o} = \sum_{j=1}^{N} \bm{\alpha}_{j} \bm{l}^{(j)}$, where $\bm{\alpha}_{j}$ represents the weight for convolved features, $\bm{l}^{(j)} \in \mathbb{R}^{K}$ is the $j$th column in $\bm{L}$, $1 \leq j \leq N$. Thus the model can focus more on significant signal locations and have better beat level interpretation. Details of \textit{knowledge-guided attention} will be introduced in Section \ref{sec:knowledge}.

\subsubsection{Rhythm Level Attentive Recurrent Layer}
For rhythm level patterns, we mainly consider the abnormal rhythm variation. 
To capture them from beat sequences, RNNs are a natural choice due to their abilities to learn on data with temporal dependencies. Again to improve interpretability and accuracies, we use \textit{knowledge guided attention} with rhythm knowledge.

Specifically, we use a bidirectional Long Short-Term Memory network~\cite{schuster1997bidirectional} (Bi-LSTM) to get rhythm level annotations of segments. The bidirectional LSTM is denoted here as  $\bm{h}^{(k)}=BiLSTM(\bm{o}^{(1)},...,\bm{o}^{(k)})$. We concatenate the forward and backward outputs of Bi-LSTM and receive the rhythm level feature $\bm{H} \in \mathbb{R}^{J \times M}$, $\bm{H}=[\bm{h}^{(1)},...,\bm{h}^{(M)}]$, $1 \leq k \leq M$.
Here we use \textit{knowledge-guided attention} with rhythm knowledge to output the rhythm level attention $\bm{c} = \sum_{k=1}^{M} \bm{\beta}_{k} \bm{h}^{(k)}$, where $\bm{\beta}_k$ represents the weight of $k$th rhythm level hidden state $\bm{h}^{(k)}$.

\subsubsection{Fusion and Prediction}
At the beginning we decompose ECG signals into multiples channels (i.e., frequency bands) and learn rhythm level features $\{\bm{c}^{(i)}\}$ from each channel $i$. Now we will perform attention fusion across all channels to have a more comprehensive view about the signal. 

We first perform fully connected transformation: $\bm{Q}=\bm{W_c}^T\bm{C}\oplus\bm{b_c}$, where $\bm{C} \in \mathbb{R}^{J \times F}$, $\bm{C}=[\bm{c}^{(1)},...,\bm{c}^{(F)}]$, $\bm{W_c} \in \mathbb{R}^{J \times H}, \bm{b_c} \in \mathbb{R}^{H}$ and $\bm{Q} \in \mathbb{R}^{H \times F}$. $\oplus$ means broadcasting $\bm{b_c}$ to all $N$ column vectors in $\bm{W_c}^T\bm{C}$ and applies addition. Then, since the importance of these channels may not be homogeneous, we will take  weighted average of $\bm{q}^{(i)}$ to calculate frequency level attention $\bm{d} = \sum_{i=1}^{F} \bm{\gamma}_i \bm{q}^{(i)}$ where $\bm{\gamma}_i$ is the weight of $\bm{q}^{(i)}$, $\bm{Q}=[\bm{q}^{(1)},...,\bm{q}^{(F)}]$, $1 \leq i \leq F$. We use frequency knowledge, signals with greater energy are more informative, to determine the weight $\bm{\gamma}$. Here we use power spectral density to measure energy.

Last, given integrated features $\bm{d}$ we make prediction using $\bm{p} = softmax(\bm{W}^T\bm{d}+\bm{b})$, where $\bm{W} \in \mathbb{R}^{H \times C}, \bm{d} \in \mathbb{R}^{H}, \bm{b} \in \mathbb{R}^{C}$, and optimize the weighted cross entropy loss $CE(\bm{p}) = - \sum_{c=1}^{C} \mathbb{I}\{\bm{z}_c=1\} \bm{w}_c \log \bm{p}_c$, where $C$ is the number of classes, $\bm{z}$ is the ground truth , $\bm{w}$ is the weight vector with the same shape as $\bm{z}$, $\mathbb{I}$ is the indication function. $\bm{w}$ is adjusted to handle with class imbalance problem which is  common in medical area. 

\subsection{Knowledge Guided Attention of \mname} \label{sec:knowledge}
We now describe how to compute multilevel attention weights $\bm{\alpha}, \bm{\beta}, \bm{\gamma}$. The attention mechanism can be regarded as a two-layer neural network: the $1$st fully connected layer calculates the scores for computing weights; the $2$nd fully connected layer computes the weights with via softmax activation.

In the first layer, the scores are computed based on the following features. 
(1) Multilevel outputs $\bm{L} \in \mathbb{R}^{K \times N}$, $\bm{H} \in \mathbb{R}^{J \times M}$, $\bm{Q} \in \mathbb{R}^{H \times F}$ extracted by \mname. (2) Domain knowledge features including beat level  $\bm{K}_{\alpha} \in \mathbb{R}^{E_{\alpha} \times N}$, rhythm level  $\bm{K}_{\beta} \in \mathbb{R}^{E_{\beta} \times M}$, and frequency level  $\bm{K}_{\gamma} \in \mathbb{R}^{E_{\gamma} \times F}$.
Concretely, three levels of domain knowledge features can be represented as below.
\begin{itemize}[leftmargin=5mm]
\itemsep0em
\item \textbf{Beat Level $\bm{K}_{\alpha}$}: For beat level knowledge we mainly consider the abnormal wave shapes or sharply changed points such as QRS complex ~\cite{kashani2005significance}. To represent it we compute first-order difference $\Delta$ and a convolutional operation $Conv_{\alpha}$ on each segment $\bm{s}$ to extract the beat level knowledge feature $\bm{K}_{\alpha} = Conv_{\alpha}(\Delta(\bm{s})$), $\Delta(\bm{s})_i = \bm{s}_i-\bm{s}_{i-1}$ and $\Delta(\bm{s})_0 = \bm{s}_0$, $\bm{s}_i$ is the $i$th value in $\bm{s}$. 
Detailed configurations of $Conv_{\alpha}$ are introduced in Implementation Details section. 
\item \textbf{Rhythm Level $\bm{K}_{\beta}$}: Attention weights $\bm{\beta}$ focus on rhythm level variation, such as severe fluctuation in ventricular fibrillation disease ~\cite{yanowitz2012introduction}. To characterize it we compute standard deviation on each segment in $\bm{S}$ to extract the rhythm level knowledge feature vector $\bm{K}_{\beta} = \mathcal{D}(\bm{S})$, where $\mathcal{D}$ calculate standard deviation of each $\bm{s}$ in $\bm{S}$, $\mathcal{D}(\bm{s}) = \frac{1}{T}\sum(\bm{s}_i - \bar{\bm{s}})^2$
\item \textbf{Frequency Level $\bm{K}_{\gamma}$}:  On frequency level, signals with greater energy contain more information and thus need more attention~\cite{yanowitz2012introduction}. So we use power spectral density (PSD), a popular measure of energy, to extract the frequency level knowledge feature vector $\bm{K}_{\gamma} = \mathcal{F}(\bm{X})$, where $\mathcal{F}$ calculate PSD \cite{Oppenheim:1996:SAS:248702} using a periodogram of each $\bm{x}^{(i)}$ in $\bm{X}$. 
\end{itemize}
Then, we concatenate model outputs and knowledge features to compute scores and attention weights. 
\begin{eqnarray*}
&\bm{\alpha} = softmax(\bm{V}_{\alpha}^T(\bm{W}_{\alpha}^T\begin{bmatrix}
\bm{L} \\
\bm{K}_{\alpha}
\end{bmatrix} \oplus \bm{b}_{\alpha})) \\
&\bm{\beta} = softmax(\bm{V}_{\beta}^T(\bm{W}_{\beta}^T\begin{bmatrix}
\bm{H} \\
\bm{K}_{\beta}
\end{bmatrix} \oplus \bm{b}_{\beta})) \\
&\bm{\gamma} = softmax(\bm{V}_{\gamma}^T(\bm{W}_{\gamma}^T\begin{bmatrix}
\bm{Q} \\
\bm{K}_{\gamma}
\end{bmatrix} \oplus \bm{b}_{\gamma}))
\end{eqnarray*}
where, $
\bm{W}_{\alpha} \in \mathbb{R}^{(K+E_{\alpha}) \times D_{\alpha}}, 
\bm{W}_{\beta} \in \mathbb{R}^{(J+E_{\beta})  \times D_{\beta}}, 
\bm{W}_{\gamma} \in \mathbb{R}^{(H+E_{\gamma}) \times D_{\gamma}}, 
\bm{b}_{\alpha} \in \mathbb{R}^{D_{\alpha}} , 
\bm{b}_{\beta} \in \mathbb{R}^{D_{\beta}}, 
\bm{b}_{\gamma} \in \mathbb{R}^{D_{\gamma}}$ 
represent weights and biases in the first layer, $\bm{V}_{\alpha} \in \mathbb{R}^{D_{\alpha} \times 1}, \bm{V}_{\beta} \in \mathbb{R}^{D_{\beta} \times 1}, \bm{V}_{\gamma} \in \mathbb{R}^{D_{\gamma} \times 1}$ represent weights in the second layer. $\oplus$ is addition with broadcasting. 

\subsection{Method for Evaluating Interpretability and Robustness}
\label{sec:interprete}

\begin{figure}[t]
\centering
\includegraphics[width=0.48\textwidth]{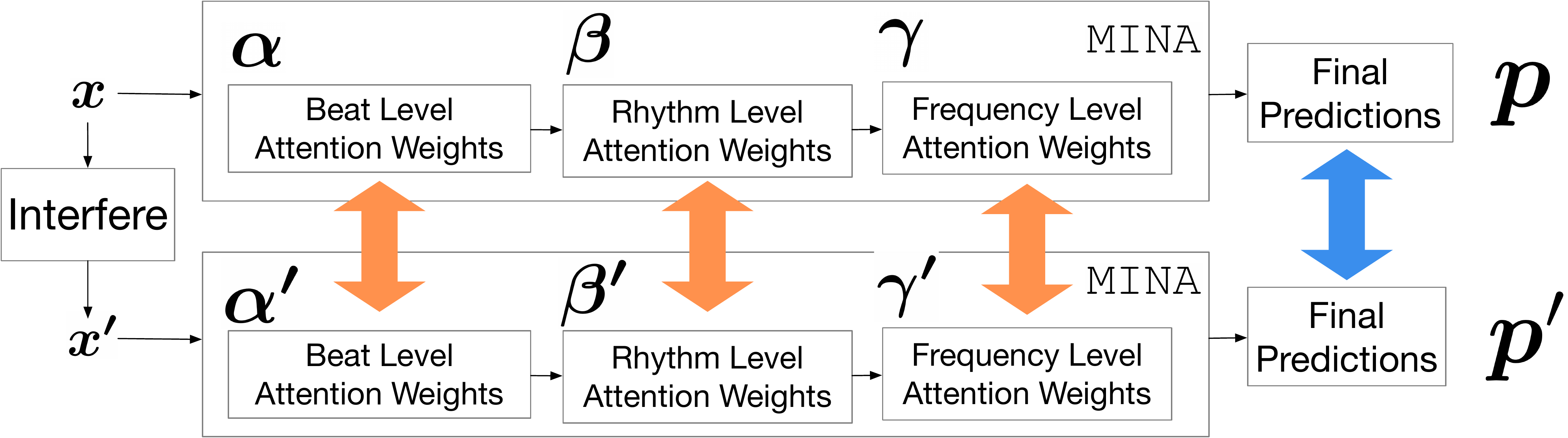}
\vspace{-0.5cm}
\caption{Analysis of multi-level attention change (Orange) and final prediction change (Blue). }
\label{fig:interp_method}
\end{figure}

To evaluate the interpretability and robustness of \mname, we perturb the signals and observe attention weights and prediction results. The evaluation method is illustrated in Figure~\ref{fig:interp_method}. 

Concretely, we add signal distortion (low frequency interferer) or noise (high frequency interferer) to the original ECG signal $\bm{x}$ and get $\bm{x'}$, here we choose baseline signal distortion and white noise. 
For the perturbed signals $\bm{x'}$, we applied \mname  to generate prediction $\bm{p'}$ and output multilevel attention weights $\bm{\alpha'}, \bm{\beta'}, \bm{\gamma'}$. We compare them with the original results $\bm{p}$ and $\bm{\alpha}, \bm{\beta}, \bm{\gamma}$ from unperturbed data. 

To evaluate the interpretability of \mname, we visually check whether attention weights are in line with medical evidences. For beat level attention weights of $M$ segments $A=[\bm{\alpha}^{(1)},...,\bm{\alpha}^{(M)}$] $ \in \mathbb{R}^{M*N}$ and $A'=[\bm{\alpha'}^{(1)},...,\bm{\alpha'}^{(M)}$] $ \in \mathbb{R}^{M*N}$, we align them to input ECG signals $\bm{x} \in \mathbb{R}^n $, where the $i$th attention weight $A_j$ approximately corresponds from $x_{\lfloor \frac{n*j}{M*N} \rfloor}$ to $x_{\lceil \frac{n*(j+1)}{M*N} \rceil}$. Then we visualize the values and verify whether high $A_j$ relates to beat level medical evidence. For rhythm level attention weight $\bm{\beta}$ and $\bm{\beta'}$, we align them to $M$ segments $\bm{S}=[\bm{s}^{(1)},...,\bm{s}^{(M)}]$, where $\bm{\beta}_k$ corresponds to $\bm{s}^{(k)}$. Then we verify whether high $\bm{\beta}_k$ relates to rhythm level medical evidence. For frequency level attention weight $\bm{\gamma}$ and $\bm{\gamma'}$, we align them to $F$ channels $\bm{X}=[\bm{x}^{(1)},...,\bm{x}^{(F)}]$, where $\bm{\gamma}_i$ corresponds to $\bm{x}^{(i)}$. Likewise, we check whether high $\bm{\gamma}_i$ relates to frequency level medical evidence.

We evaluate the robustness of \mname based on the two tasks: (1) we visually compare whether the new attention weights after perturbation are still in line with medical evidences, using the same way above, (2) we gradually change the interfered amplitude and evaluate the overall performance changes. The more robust model will be less impacted. Moreover, these results can also be used to evaluate interpretability, since interpretable model can highlight meaningful information, while also suppress unrelated parts. 

\section{Experiments}
\label{sec:experiments}

In this section, we first describe the dataset used for the experiments, followed by the description of the baseline models. Then we discuss the model performance.

\subsection{Source of Data}

We conducted all experiments using real world ECG data from PhysioNet Challenge 2017 databases ~\cite{clifford2017af}. The dataset contains 8,528 de-identified ECG recordings lasting from 9s to just over 60s and sampled at 300Hz by the AliveCor device, 738 from AF patients and 7790 from controls as predefined by the challenge. We first divided the data into a training set (75\%), a validation set (10\%) and a test set (15\%) to train and evaluate in all tasks. Then, we preprocess them to get equal length data, where $n=3000$. The summary statistics of the data is described in Table~\ref{tb:summary}. In this study, the objective is to discriminate records of AF patients from those of controls. 

\begin{table}[t]
\centering
\resizebox{0.48\textwidth}{!}{
\begin{tabular}{llccccc} 
\hline
\multirow{2}{*}{Type} & \multirow{2}{*}{\# recording} & \multicolumn{5}{c}{\# of points}                                                                                                               \\ 
\cline{3-3}\cline{4-4}\cline{5-5}\cline{6-6}\cline{7-7}
                      &                               & \multicolumn{1}{c}{Mean}         & \multicolumn{1}{c}{StDev} & \multicolumn{1}{c}{Max}& \multicolumn{1}{c}{Median}& \multicolumn{1}{c}{Min}  \\ 
\hline
AF                    & \multicolumn{1}{c}{738}      & 9631                              & 3703                     & 18062& 9000& 2996                                                                \\ 
non-AF                & \multicolumn{1}{c}{7790}     & 9760                              & 3222                     & 18286& 9000& 2714                                                                \\
\hline
\end{tabular}
}
\caption{Data profile of PhysioNet Challenge 2017 dataset}
\label{tb:summary}
\end{table}

\subsection{Baseline Models}

We will compare \mname with the following models:
\begin{itemize}
\item \textbf{Expert}: A combination of extracted features used in AF diagnosis including: rhythm features like sample entropy on QRS interval \cite{expert_1}; cumulative distribution functions \cite{tateno2001automatic}; thresholding on the median absolute deviation (MAD) of RR intervals \cite{expert_3}; heart rate variability in Poincare plot \cite{park2009atrial}; morphological features like location, area, duration, interval, amplitude and slope of related P wave, QRS complex, ST segment and T wave; frequency features like frequency band power. We used QRS segmentation method in ~\cite{pan1985real} and trained an LR classifier using these features. Then, we build both logistic regression (\textbf{ExpertLR}) and random forest (\textbf{ExpertRF}) on above extracted features. 
\item \textbf{CNN}: Convolutional layers are performed on ECG segments with shared weights. We use global average pooling to combine features, and fully connect (FC) layer and softmax layer for prediction. The model architecture is modified based on ~\cite{tbe} to handle ECG segments. The hyper-parameters in CNN, FC and softmax are the same as \mname to match the model complexity. 
\item \textbf{CRNN}: We used shared weights convolutional layers on ECG segments, 
and replaced the global average pooling with bi-directional LSTM. Then FC and softmax are applied to the top hidden layer. The architecture is modified based on ~\cite{zihlmann2017convolutional}, but only keep one convolutional layer. Other hyper-parameters in CNN, RNN, FC and softmax are the same as \mname. 
\item \textbf{ACRNN}: Based on CRNN, with additional beat level attentions and rhythm level attentions. Other hyper-parameters are the same as \mname. 
\end{itemize}

\subsection{Implementation Details}

In convolutional layers of CNN, CRNN, ACRNN and \mname, we use one layer for each model. The number of filters is set to 64, the filter size is set to 32 and strider is set to 2. Pooling is replaced by attention mechanism. $Conv_\alpha$ of $\bm{K}_{\alpha}$ has one filter with size set to 32, the strider is also 2. In recurrent layers of CRNN, ACRNN and \mname, we also use one single layer for each model, the number of hidden units in each LSTM is set to 32. The dropout rate in the fully connected prediction layer is set to 0.5. In sliding window segmentation, we use non-overlapping stride with $T=50$.
Deep models are trained with the mini-batch of 128 samples for 50 iterations, which was a sufficient number of iterations for achieving the best performance for the classification task. 
The final model was selected using early stopping criteria on validation set. We then tested each model for 5 times using different random seeds, and report their mean values with standard deviation.
All models were implemented in PyTorch version 0.3.1, and trained with a system equipped with 64GB RAM, 12 Intel Core i7-6850K 3.60GHz CPUs and Nvidia GeForce GTX 1080. All models were optimized using Adam \cite{adam}, with the learning rate set to 0.003. 
Our code is publicly available at \url{https://github.com/hsd1503/MINA}.

\subsection{Performance Comparison}

Performance was measured by the Area under the Receiver Operating Characteristic (ROC-AUC), Area under the Precision-Recall Curve (PR-AUC) and the
F1 score. The PR-AUC is considered a better measure for imbalanced data like ours~\cite{davis2006relationship}. Table~\ref{table:compare} shows \mname outperforms all baselines, and shows $5.51\%$ higher PR-AUC than the second best models.

\begin{table}[t]
\centering
\resizebox{0.48\textwidth}{!}{
\begin{tabular}{llll}
\hline
  & ROC-AUC    & PR-AUC     & F1    \\ \hline
ExpertLR & 0.9350 $\pm$ 0.0000 & 0.8730 $\pm$ 0.0000 & 0.8023 $\pm$ 0.0000  \\
ExpertRF & 0.9394 $\pm$ 0.0000 & 0.8816 $\pm$ 0.0000 & 0.8180 $\pm$ 0.0000  \\
CNN & 0.8711 $\pm$ 0.0036 & 0.8669 $\pm$ 0.0068 & 0.7914 $\pm$ 0.0090  \\ 
CRNN & 0.9040 $\pm$ 0.0115 & 0.8943 $\pm$ 0.0111 & 0.8262 $\pm$ 0.0215 \\ 
ACRNN & 0.9072 $\pm$ 0.0047 & 0.8935 $\pm$ 0.0087 & 0.8248 $\pm$ 0.0229  \\ \hline
\textbf{\mname} & \textbf{0.9488} $\pm$ 0.0081 & \textbf{0.9436} $\pm$ 0.0082 & \textbf{0.8342} $\pm$ 0.0352 \\ \hline
\end{tabular}}
\caption{Performance Comparison on AF Prediction}
\vspace{-0.05in}
\label{table:compare}
\end{table}

\section{Interpretability and Robustness Analysis}

\subsection{\mname Automatically Extracts Clinically Meaningful Patterns} 

\begin{figure}[t]
\centering
\includegraphics[width=0.48\textwidth]{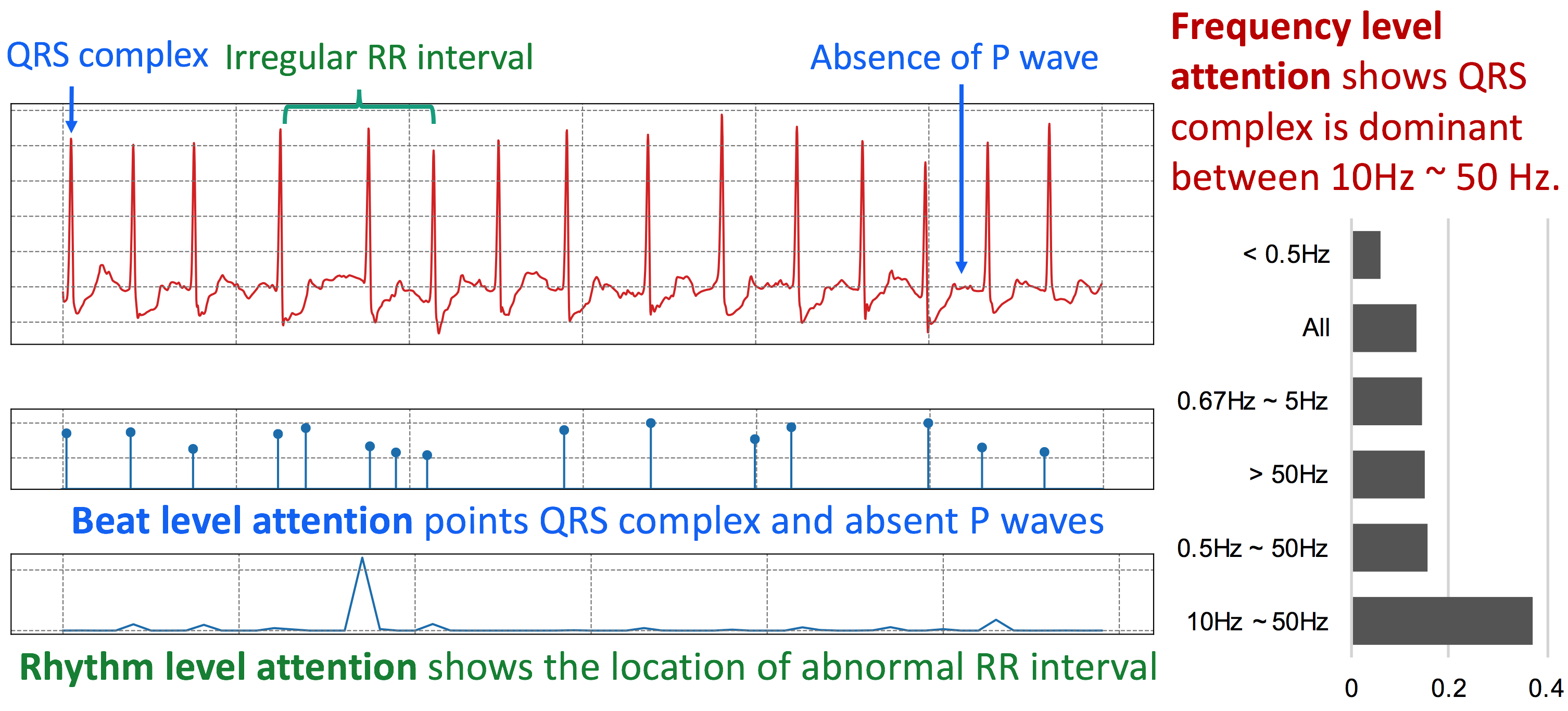}
\caption{An ECG signal of AF patient (left top), \mname learns beat level attention which points to the position of significant QRS complexes and abnormal P waves. Rhythm level attention shows the abnormal RR interval. The frequency channel with highest attention correspond to the frequency bands where QRS complex is dominant. }
\label{fig:interpretation_example}
\end{figure}

When reading an ECG record (upper left in Figure~\ref{fig:interpretation_example}), cardiologists will make AF diagnosis based on following clinical evidences: 1) the absence of P wave: a small upward wave before QRS complex; 2) the irregular RR interval such as the much wider one between the $4$th and the $5$th QRS complex.

\mname learns these patterns automatically via beat-, rhythm-, and frequency level attention weights. From Figure~\ref{fig:interpretation_example}, the beat level attentions point to where QRS complex or absent P waves occur. The rhythm level attentions indicate the location of abnormal RR interval, which 
precisely matches the clinical evidence. Besides, from the frequency level attentions, we notice channel 10Hz-50Hz receives the highest attention weight so \mname pays more attention to it. In fact, QRS complex, the most significant clinical evidence in ECG diagnosis, is known to be dominant in 10Hz-50Hz~\cite{tateno2001automatic,expert_3,expert_1}. 

\subsection{\mname Remains Interpretable and Robust Against Baseline Signal Distortion}

The baseline wander distortion is a low frequency noise with slow but large changes of the signal offset. It is a common issue that drops ECG analysis performance. In this experiment, we mimic the real world setting by distorting data and observe whether \mname can still make robust and interpretable predictions.

\begin{figure}[t]
\centering
\includegraphics[width=0.48\textwidth]{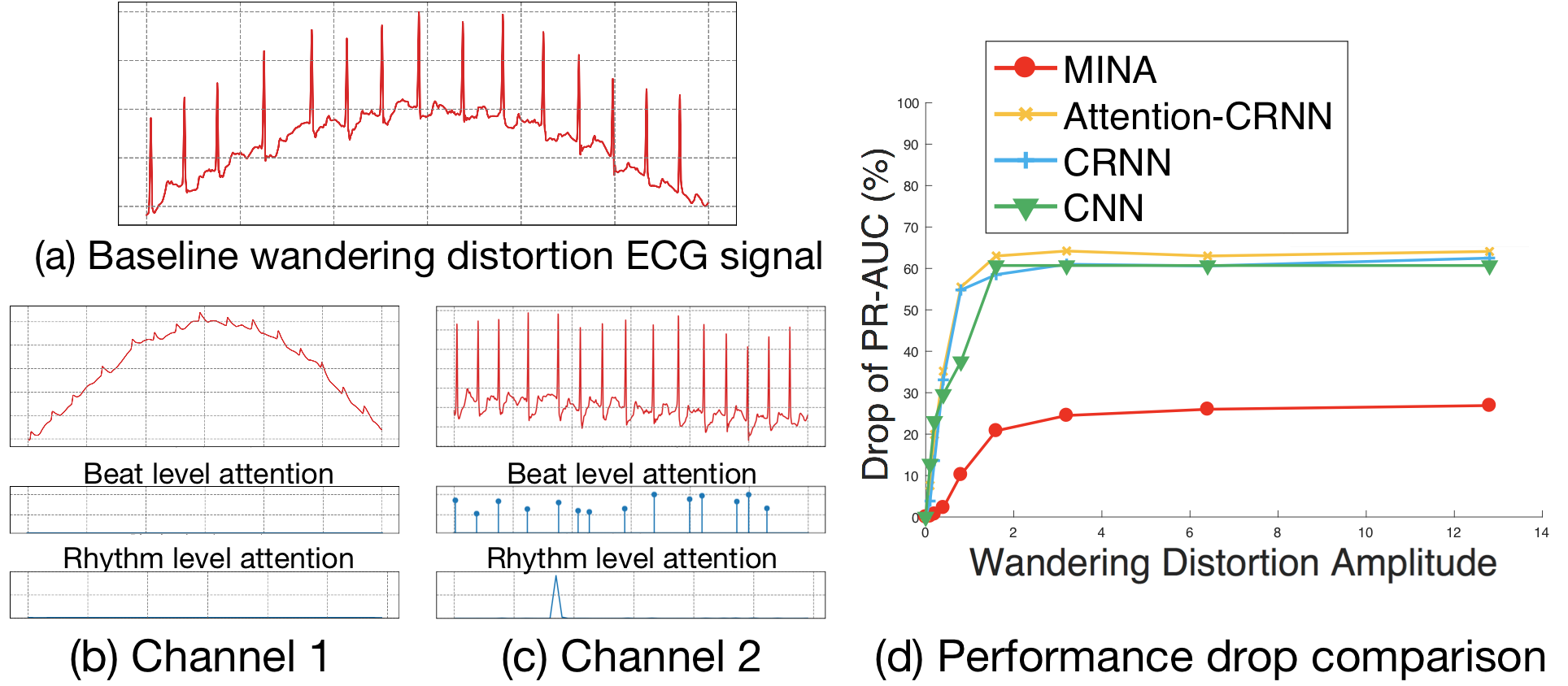}
\caption{(a) Signal in Figure \ref{fig:interpretation_example} interfered by baseline wander distortion. (b) Channel 1 (low attention weights) shows no significant patterns. (c) Channel 2 (higher attention weights) remains meaningful patterns similar to original data. (d) \mname has much lower PR-AUC drop $\%$ than baselines.}
\label{fig:wander}
\end{figure}

For the experiment we interfered the signal in Figure~\ref{fig:interpretation_example} with baseline wander distortion. The interfered signal is plotted in Figure~\ref{fig:wander}(a). From the original frequency attention in Figure~\ref{fig:interpretation_example}, it is easy to see Channel 1 ($<$0.5Hz) has the lowest weights, while Channel 2 (0.5Hz-50Hz) weights much higher. Thus Channel 1 can be interpreted as baseline component while Channel 2 as clean signal component. \mname pays more attention to Channel 2 than Channel 1. After signal distortion, the importance of both channels remain the same, which is also reflected from their beat level and rhythm level attentions. 
Channel 1 shows no significant patterns, but the more informative Channel 2 have similar beat- and rhythm level patterns as unperturbed data, which indicates the interpretability of  \mname will be less impacted by data distortion.

To evaluate model robustness, we compare the performance change along the increase of distortion amplitude on the entire test set. As shown in Figure \ref{fig:wander}(d), \mname still has much lower performance drop even after distortion by large amplitude. While all baselines start to have large performance drop even with little distortion. This is mainly thanks to frequency attention fusion. In training process, the model already identified  Channel 1 a baseline signal. Thus baseline distortion will have less impact on important signals in clean signal channel. Since baseline signal distortion occurs in real clinical setting, \mname will provide more accurate prediction in these scenarios. 

\subsection{\mname Remains Interpretable and Robust in the Presence of Noise}

\begin{figure}[t]
\centering
\includegraphics[width=0.48\textwidth]{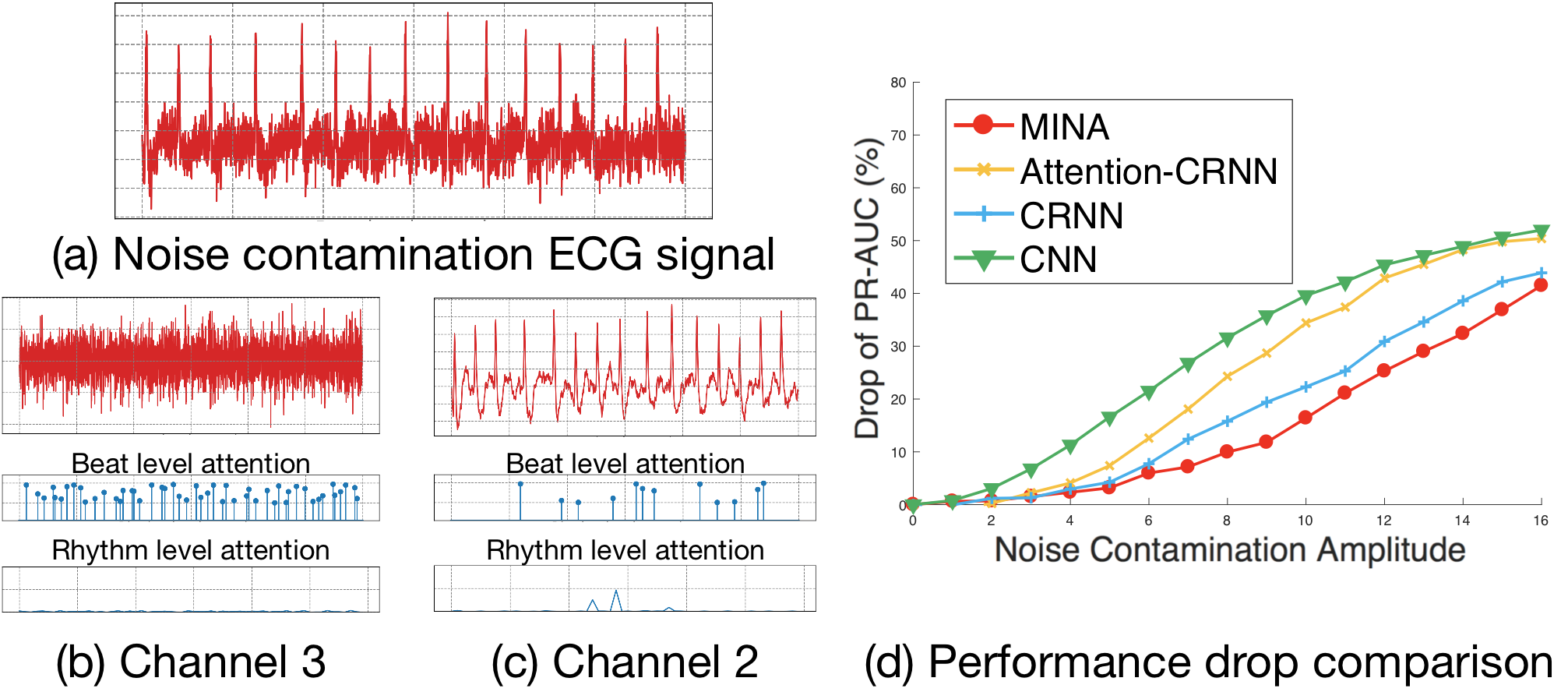}
\caption{(a) Signal in Figure \ref{fig:interpretation_example} perturbed by noise. (b) Channel 3 (lower attention weights) shows no significant patterns. (c) Channel 2 (higher attention weights) remains meaningful patterns similar to original data. (d) \mname has much lower PR-AUC drop $\%$ than baselines.}
\label{fig:noise}
\end{figure}

The high frequency noise contamination is another common issues. For this experiment, we perturbed the signal in Figure~\ref{fig:interpretation_example} with white noise. The perturbed signal is in Figure~\ref{fig:noise}(a). Similar to last experiment, from original frequency attentions we know Channel 3 ($>$50Hz) has lower weights. It is a channel known for high noise. While Channel 2 (0.5Hz-50Hz) weights much higher and is known as a clean signal channel. 

After noise contamination, the noise impacts more to the noise Channel which is less important in the prediction of \mname, but the more informative Channel 2 have similar beat- and rhythm level patterns as unperturbed data, which indicates the interpretability of \mname will be less impacted by noise contamination. In Figure~\ref{fig:noise}(d), we compare the PR-AUC change along the increase of noise amplitude on the entire test set.  \mname is less impacted by noise than other methods, demonstrating more robust performance in the presence of noise thanks to frequency attention fusion.

\section{Conclusion and Future Work}

In this paper, we propose \mname, a deep multilevel knowledge-guided attention networks that interpretatively predict heart disease from ECG signals. 
\mname outperformed baselines in heart disease prediction task. Experimental results also showed robustness and strong interpretability against signal distortion and noise contamination. In future, we can extend to a broad range of disease where ECG signals can be treated as additional information in the diagnosis, on top of other health data such as electronic health records. 
Then we will need to investigate interpretable prediction based on multimodal data, which is a possibly rewarding avenue of future research.

\section*{Acknowledgements}
This work was supported by the National Science Foundation, award IIS-1418511, CCF-1533768 and IIS-1838042, the National Institute of Health award 1R01MD011682-01 and R56HL138415.
We also thanks valuable discussions with Li Jiang from BOE.

\bibliographystyle{named}
\bibliography{reference}

\ifthenelse{\value{sol}=1}{

\clearpage

\appendix{\mname: Multilevel Knowledge-Guided Attention for Modeling Electrocardiography Signals}

\section{Background of Electrocardiography (ECG)}\label{app:ecg}

The Electrocardiography (ECG) is a test that measures the electrical activity of the heartbeat. With each beat, an electrical impulse travels through the heart and causes the muscle to squeeze and pump blood from the heart. Then ECG signals will record the timing of the top and lower chambers.

A normal heart beat in ECG is shown in Figure \ref{fig:ecg}. Usually a "P wave" which is characterized by the right and left atria or upper chambers will arrive first, following by a flat line indicating when electrical impulse goes to the bottom chambers. Then next wave called ventricular depolarization (QRS complex) arrive. 
The next wave is called ventricular repolarization (ST segment, T wave), which represents electrical recovery or return to a resting state for the ventricles. Together we also have "U wave" that represents papillary muscle repolarization.

\begin{figure}[H]
\centering
\includegraphics[width=0.33\textwidth]{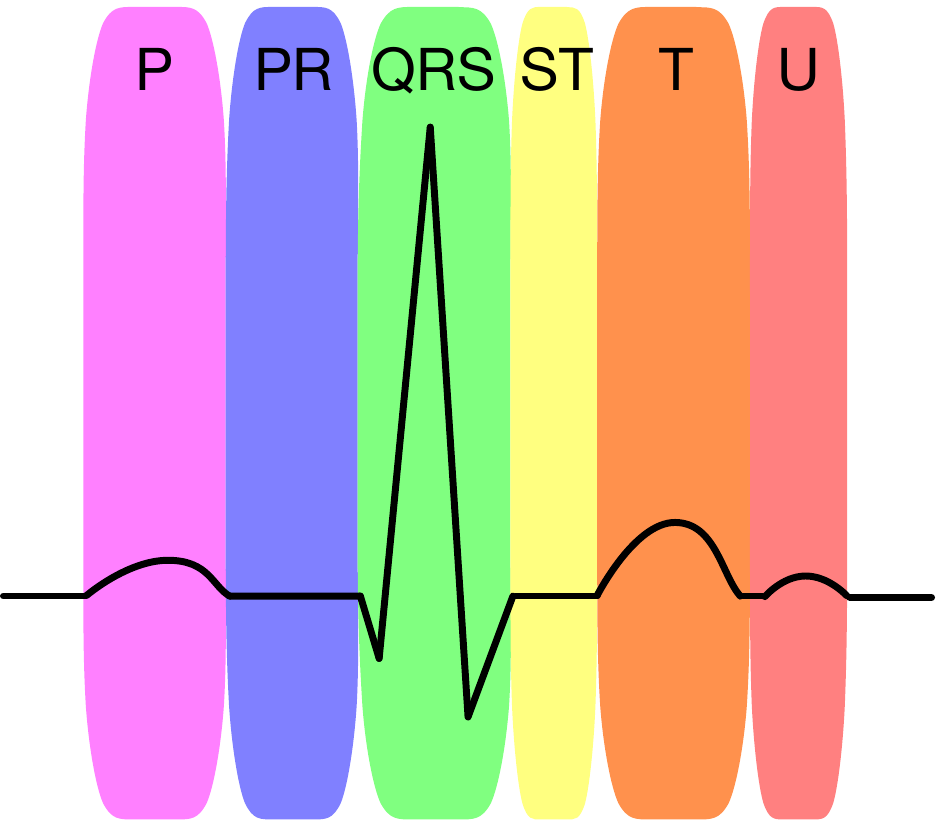}
\caption{A normal heart beat. }
\label{fig:ecg}
\end{figure}

ECG signals offer two types of information: 1) the \textit{time interval}s measures how long the electrical wave needs to pass through the heart: normal or slow, fast or irregular; 2) the \textit{amount} of electrical activity passing through the heart shows whether the size of parts of the heart become abnormal.

The time domain features for heart disease diagnosis include beat level and  rhythm level. 
\begin{itemize}[leftmargin=5mm]
\itemsep0em
\item In beat level, an unusual p-wave may indicate disease such as atrial fibrillation (AF), ectopic atrial pacemaker, atrial enlargement et al. An unusual QRS complex may indicate disease such as left/right bundle branch block and ventricular tachycardia. An unusual ST segment and T wave may indicate myocardial infarction, ischemia, and left ventricular hypertrophy.
\item In rhythm level, the analysis is usually based on intervals between QRS complexes, which is called RR interval. Long RR interval may indicate sinus bradycardia, short RR interval may indicate sinus tachycardia or ventricular tachycardia, while irregular RR interval may indicate AF. However, many disease such as AF poses patterns in both beat level and rhythm level, so it is beneficial to combine them together for analysis. 
\end{itemize}

\section{Frequency Band for ECG Signals} \label{app:freq}

The ECG signal is a mixture of heart muscle's electrophysiologic activities including atrial, ventricular, papillary muscle and myocardium. Besides, it may also contain other electrical components from muscle, skin, respiration, body moving etc. The frequency bands listed below are commonly considered dominant components in ECG signal: 

\begin{itemize}[leftmargin=5mm]
\itemsep0em
\item < 0.5 Hz: very low frequency component, mainly represent heart unrelated wandering. 
\item 0.12 Hz - 0.5 Hz: respiration.
\item 0.5 Hz - 50 Hz: P wave, QRS complex and T wave. 
\item 0.67 Hz - 5 Hz: P wave. 
\item 1 Hz - 7 Hz: T wave.
\item 5 Hz - 50 Hz: muscle.
\item 10 Hz - 50 Hz: QRS complex is the most dominate component. 
\item > 50 Hz: high frequency noise. 
\item All: raw signal. 
\end{itemize}

Notice that these frequency bands are approximate, since they are hard to be divided entirely. Besides, their significance may also vary among people. However, it is beneficial to combine frequency domain features and time domain features together for disease diagnosis, since the transformation of frequency bands will divide time domain ECG signals into subspaces, thus helps classification tasks. 

The illustration of frequency transformation is shown in Figure \ref{fig:bandpass}. 

\begin{figure}[H]
\centering
\includegraphics[width=0.5\textwidth]{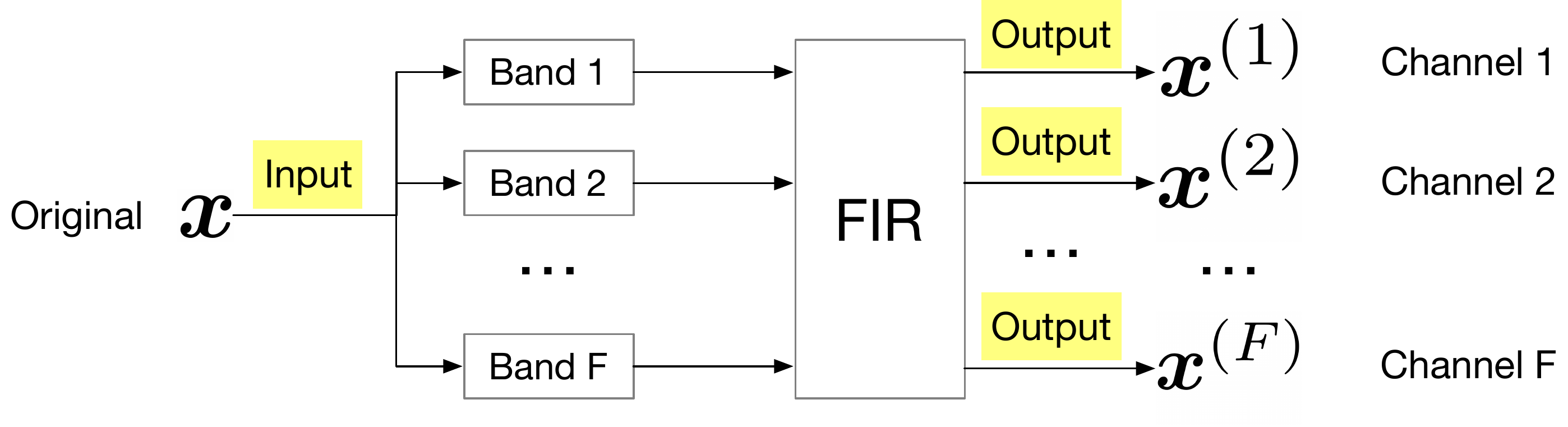}
\caption{Finite impulse response bandpass filter in frequency transformation layer. }
\label{fig:bandpass}
\end{figure}

\section{Interferer Simulation Details} \label{app:interferer}

We simulate baseline wander distortion signal using sine function and noise contamination signal using random normal distribution. Concretely, when interfere length $n$ signal $\bm{x}$: 
\begin{equation}
\begin{aligned}
\bm{x'} & =\bm{x} + amp*sin([\frac{1*\pi}{n},\frac{2*\pi}{n},...,\pi]) \\
\bm{x'} & =\bm{x} + amp*random\_normal([1,2,...,n])
\end{aligned}
\end{equation}
where $amp$ is amplitude of interfere, $+$ represents elementwise addition. 

\section{More Interpretability Evaluation Examples} \label{app:example}

\begin{figure}[H]
\centering
\begin{tabular}{c}
\includegraphics[width=0.3\textwidth]{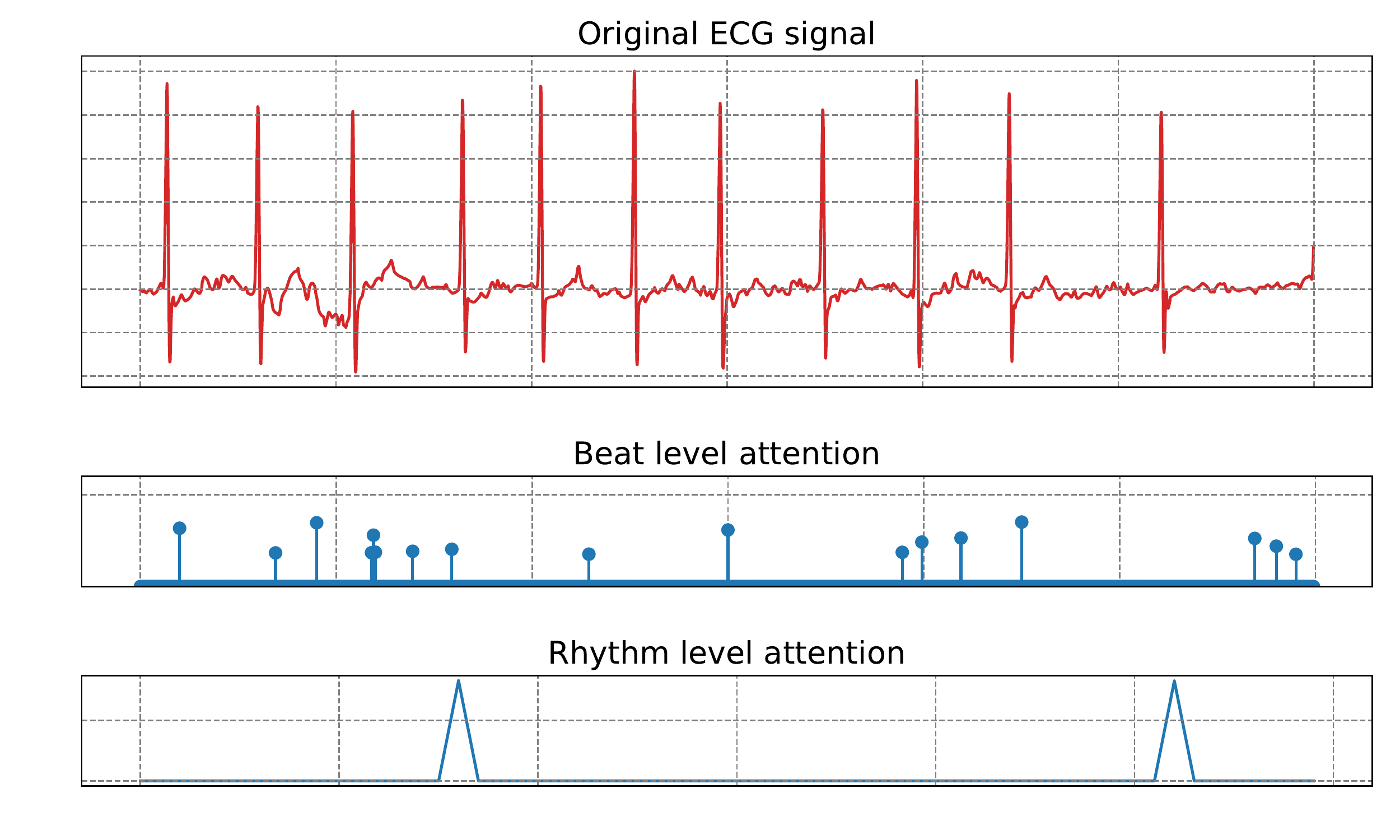}\\
(a)\\
\includegraphics[width=0.3\textwidth]{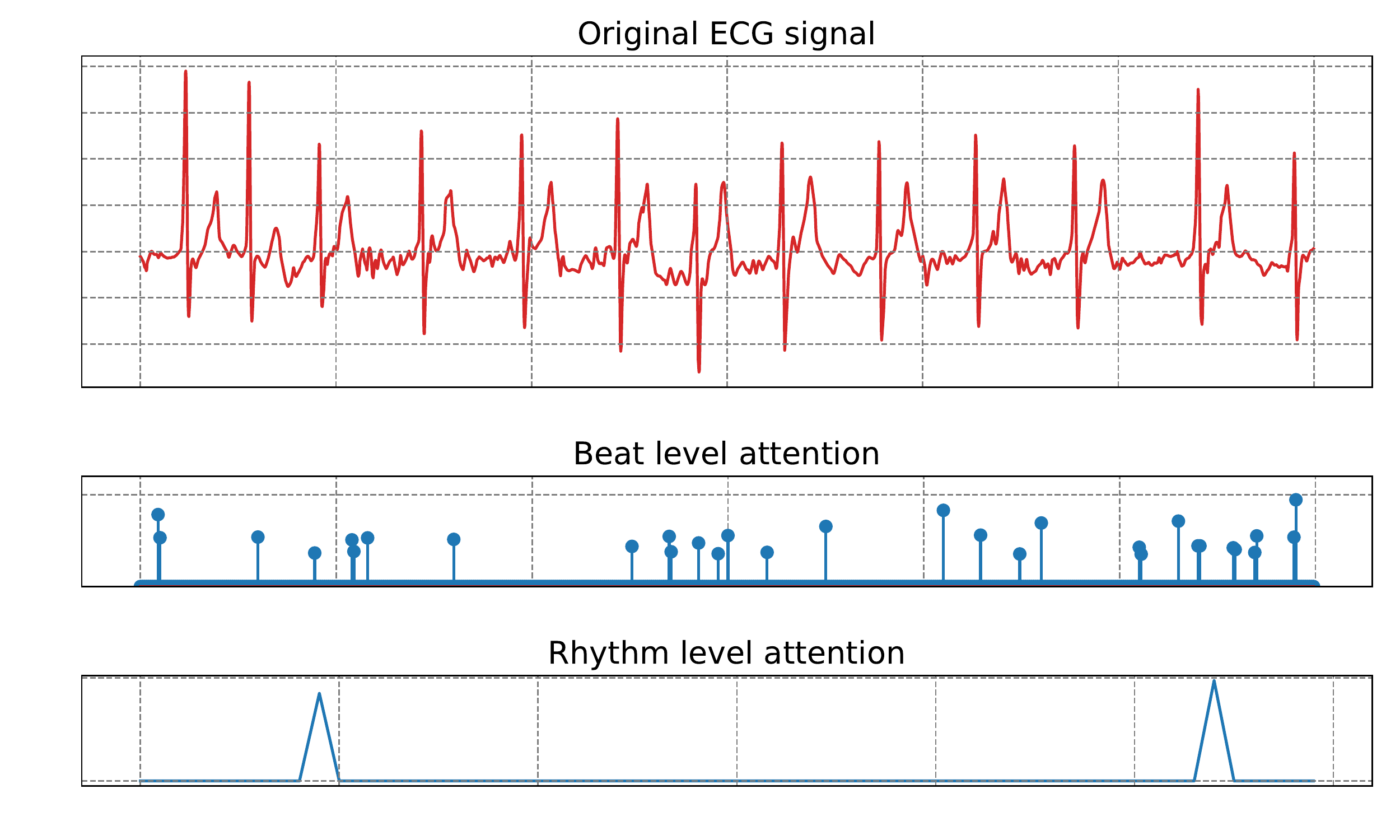}\\
(b)\\
\includegraphics[width=0.3\textwidth]{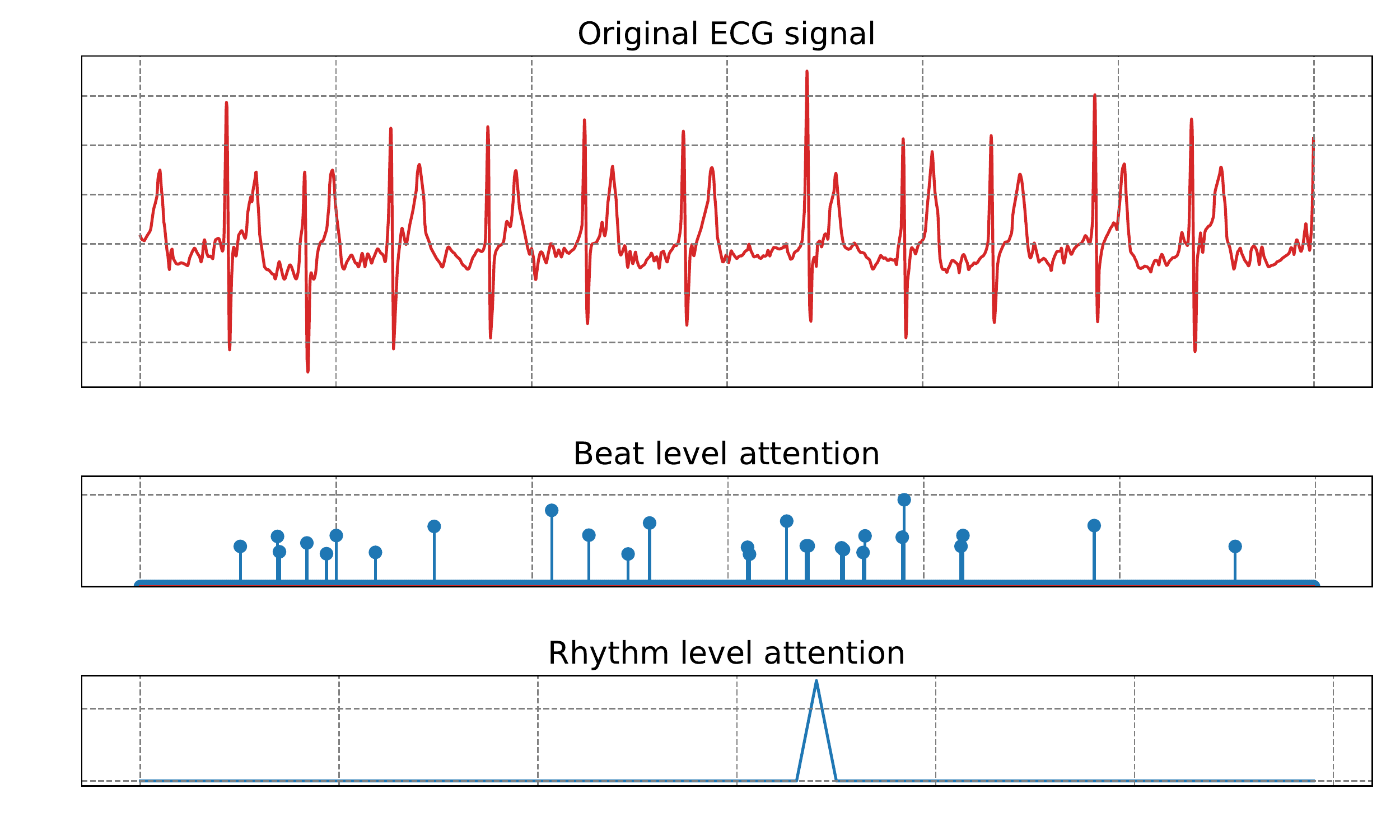}\\
(c)\\
\includegraphics[width=0.3\textwidth]{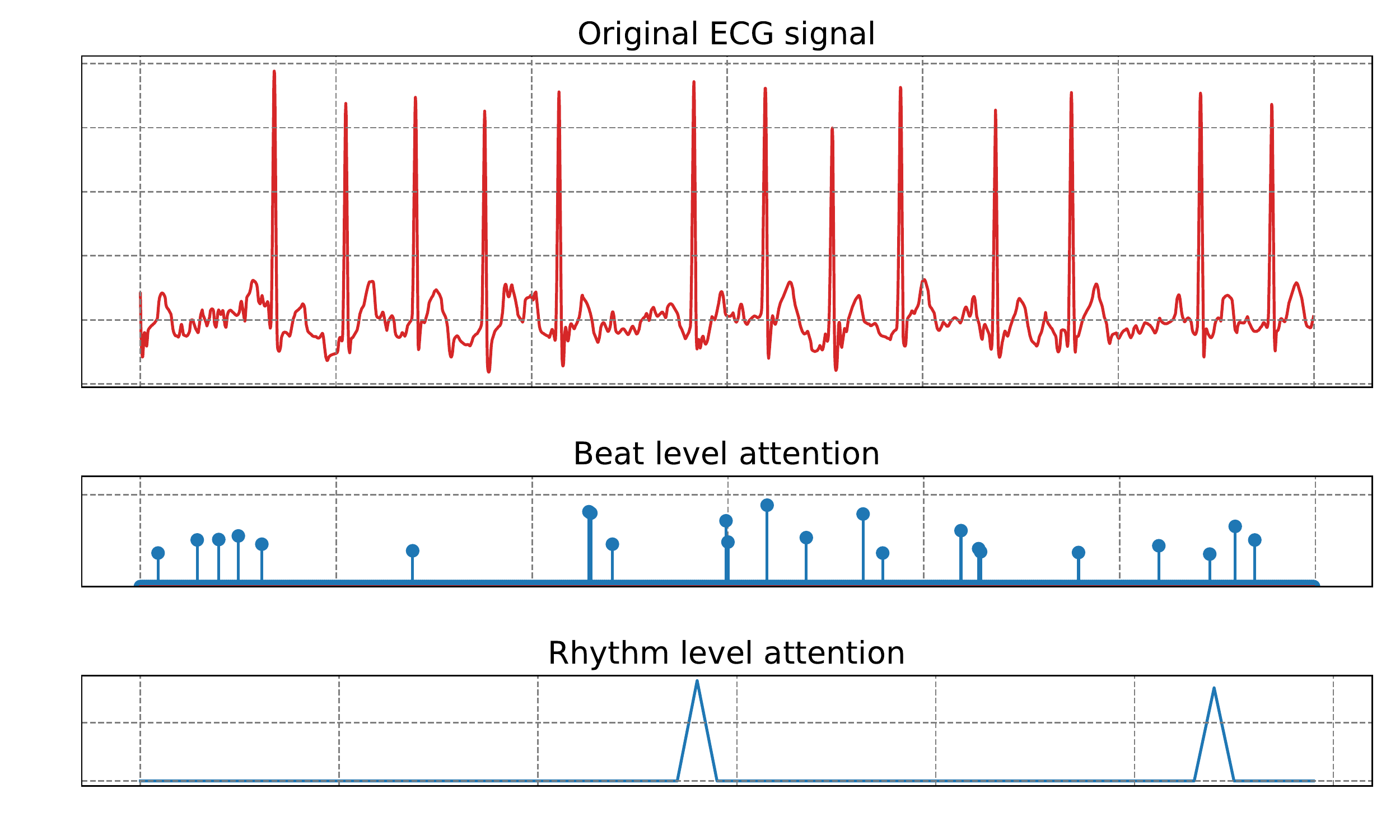}\\
(d)\\
\includegraphics[width=0.3\textwidth]{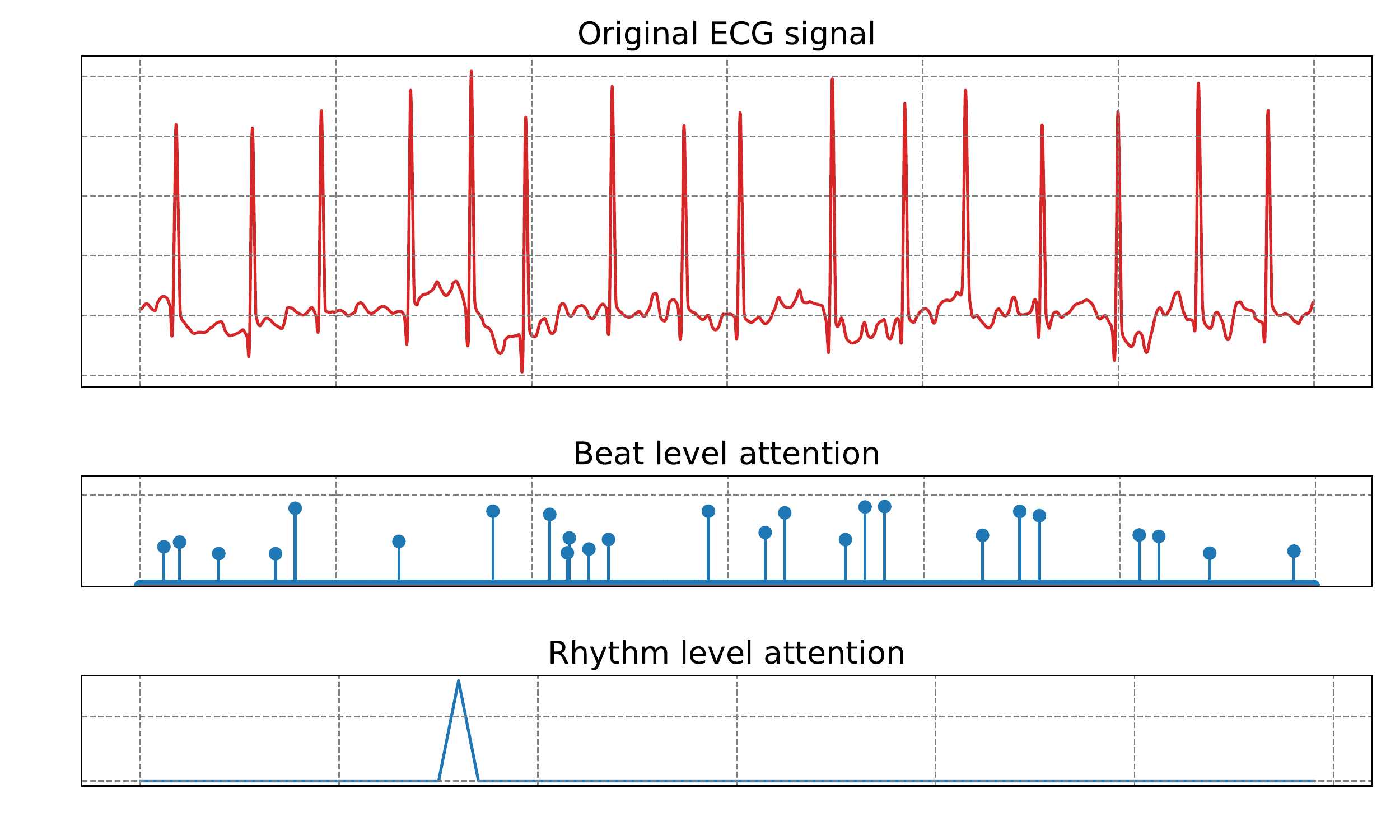}\\
(e)\\
\includegraphics[width=0.3\textwidth]{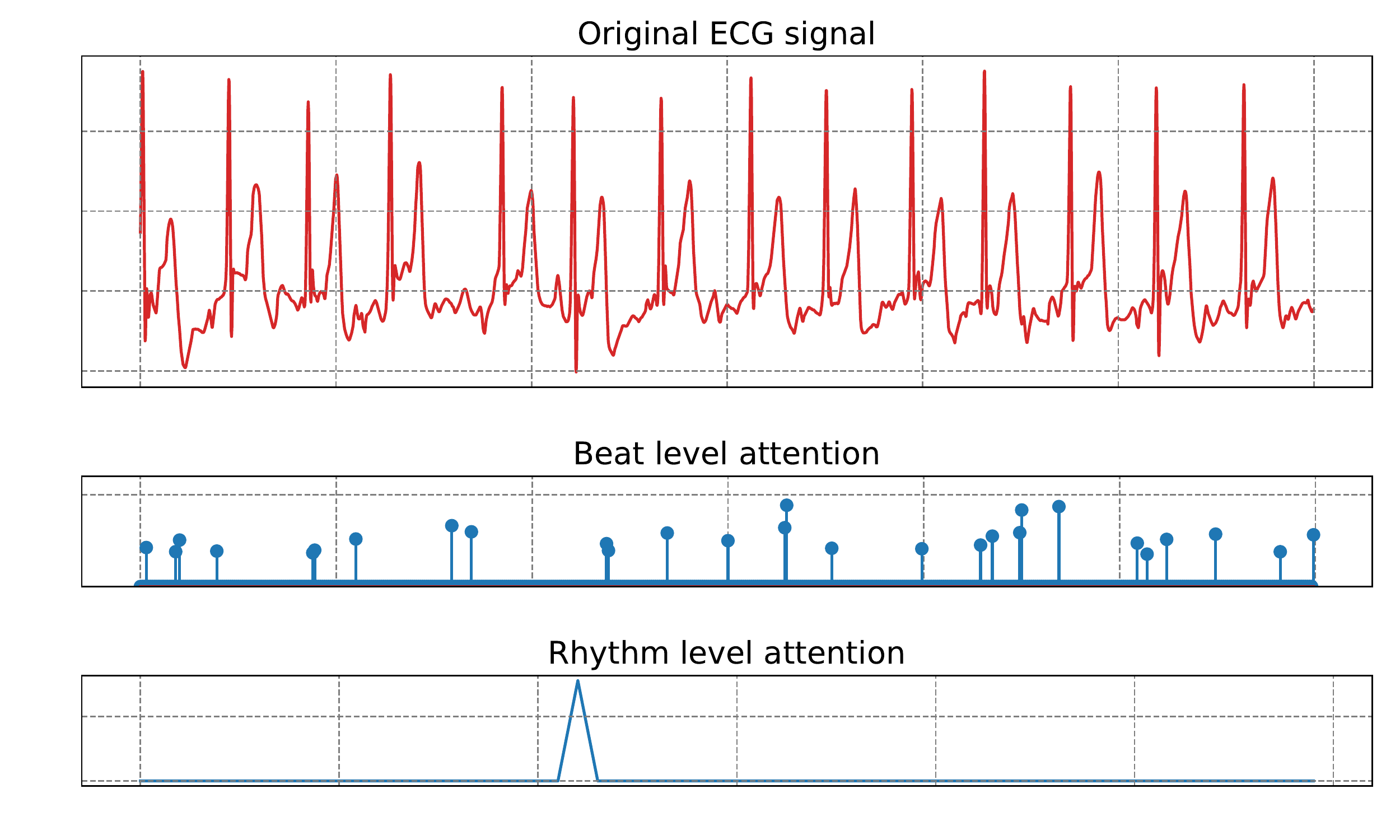}\\
(f)\\
\end{tabular}
\caption{More examples of interpretability evaluations}
\label{fig:more}
\end{figure}

}

\end{document}